\documentclass[a4paper,11pt]{article}

\pdfoutput=1

\usepackage{jheppub}
\usepackage{amsmath}
\usepackage{xspace}
\usepackage{hyperref}
\usepackage{mdwlist}

\newcommand{\nnu}{\nonumber\\}
\newcommand{\nn}{\nonumber}
\newcommand{\bef}{\begin{figure}[t]\centering}
\newcommand{\eef}{\end{figure}}
\def\bea#1\eea{\begin{align}#1\end{align}}
\def \be  {\begin{equation}}
\def \ee  {\end{equation}}
\def \ba  {\begin{eqnarray}}
\def \ea  {\end{eqnarray}}

\newcommand{\f}{\frac}

\newcommand{\GG}{{\cal G}}

\newcommand\as{\alpha_s}

\allowdisplaybreaks[2]

\title{Hadron-in-jet production at partonic threshold}

\author[a]{Tom Kaufmann,}
\author[b]{Xiaohui Liu,}
\author[c]{Asmita Mukherjee,}
\author[d,e]{Felix Ringer,}
\author[a]{Werner Vogelsang}

\affiliation[a]{Institute for Theoretical Physics, T\"ubingen University, Auf der Morgenstelle 14, 72076
T\"ubingen, Germany}
\affiliation[b]{Center of Advanced Quantum Studies, Department of Physics, Beijing Normal University, Beijing
100875, China}
\affiliation[c]{Department of Physics, Indian Institute of Technology Bombay,
Powai, Mumbai 400076, India}
\affiliation[d]{Physics Department, University of California, Berkeley, CA 94720, USA}
\affiliation[e]{Nuclear Science Division, Lawrence Berkeley National Laboratory, Berkeley, CA 94720, USA}
                                         
\emailAdd{tom.kaufmann@uni-tuebingen.de}
\emailAdd{xiliu@bnu.edu.cn}
\emailAdd{asmita@phy.iitb.ac.in}
\emailAdd{fmringer@berkeley.edu}
\emailAdd{werner.vogelsang@uni-tuebingen.de}

\abstract{We consider the longitudinal momentum distribution of hadrons inside jets in proton-proton collisions. At partonic threshold large double logarithmic corrections arise which need to be resummed to all orders. We develop a factorization formalism within SCET that allows for the joint resummation of threshold and jet radius logarithms. We achieve next-to-leading logarithmic NLL$'$ accuracy by including non-global logarithms in the leading-color approximation. Overall, we find that the threshold resummation leads to a sizable enhancement of the cross section and a reduced QCD scale dependence, suggesting that the all-order resummation can be important for the reliable extraction of fragmentation functions in global analyses when jet substructure data is included.}

\begin{document}
\maketitle

\section{Introduction}
\label{sec:intro}

In recent years jet substructure techniques have become important tools at the LHC and RHIC. The measurement of jet substructure cross sections allows for new high precision studies in QCD such as the extraction of fragmentation functions, nonperturbative shape functions and the determination of the QCD strong coupling constant. In addition, jets can be used as probes of the Quark Gluon Plasma (QGP) in heavy-ion collisions and for searches of physics beyond the standard model. See~\cite{Larkoski:2017jix} for a recent review of jet substructure techniques and their applications. In this work, we consider the longitudinal momentum distribution of hadrons inside an identified jet. The cross sections of this observable are typically reported in terms of the momentum fraction $z_h=p_T^h/p_T$, where $p_T^h$ ($p_T$) is the hadron (jet) transverse momentum measured with respect to the beam axis. In the kinematic region where the momentum fraction of the hadron relative to the jet approaches unity $z_h\to 1$, threshold logarithms in the perturbative expansion of the cross section become large and need to be resummed to all orders in QCD~\cite{Sterman:1986aj,Catani:1989ne}. The experimental Collaborations at the LHC and RHIC have reported data with $z_h$ values up to 0.8-0.9 which is very close to the hadronic threshold. Further away from the hadronic threshold, the resummation of threshold logarithms can also be important since the logarithms appear at the parton level and the perturbative parton level result is convolved with steeply falling fragmentation functions.

The longitudinal momentum distribution of hadrons inside jets was first studied in exclusive jet production~\cite{Procura:2009vm,Jain:2011xz,Jain:2011iu}. At proton-proton colliders jet substructure measurements are often performed on an inclusive jet sample $pp\to {\rm jet}+X$. The relevant calculations for the process $pp\to ({\rm jet}+h)+X$ at next-to-leading order (NLO) were carried out numerically in~\cite{Arleo:2013tya} and analytically in the narrow jet approximation in~\cite{Kaufmann:2015hma,Kang:2016ehg,Dai:2016hzf,Kang:2019ahe}, where also the resummation of logarithms of the jet radius $\alpha_s^n\ln^n R^2$ was taken into account. The longitudinal momentum distribution inside jets provides important constraints for the extraction of collinear fragmentation functions. First, the variable $z_h$ allows for a more direct ``scan'' of the fragmentation functions compared to inclusive hadron production $pp\to h+X$. Second, gluons contribute at leading-order such that the cross section can provide stringent constraints for the extraction of the gluon fragmentation function. Instead, processes such as $e^+e^-\to h+X$ are only sensitive to the gluon fragmentation function at NLO and through evolution effects. See~\cite{Metz:2016swz} for an overview of fragmentation functions and recent extractions can be found in~\cite{deFlorian:2014xna,Hirai:2016loo,Fickinger:2016rfd,deFlorian:2017lwf,Anderle:2017cgl,Bertone:2017tyb,Ethier:2017zbq,Bertone:2018ecm,Soleymaninia:2018uiv,Sato:2019yez}. By now, a wide range of identified particles inside the jet have been considered. Besides light charged hadrons, calculations have been performed for heavy-flavor mesons~\cite{Chien:2015ctp,Bain:2016clc}, quarkonium~\cite{Bain:2017wvk,Kang:2017yde} and photons~\cite{Kaufmann:2016nux}. A first global analysis of $D^*$-meson fragmentation functions including jet substructure data was performed in~\cite{Anderle:2017cgl}. Experimental results in proton-proton and heavy-ion collisions at the LHC can be found in~\cite{Aad:2011td,Aad:2011sc,Chatrchyan:2012gw,Chatrchyan:2014ava,Aad:2014wha,ATLAS:2015mla,Aaboud:2017tke,Aaij:2017fak,Acharya:2019zup} and at RHIC in~\cite{Adamczyk:2017wld,Adamczyk:2017ynk}. In this work, we extend the theoretical description of the longitudinal momentum measurement to the threshold region. We develop a joint resummation framework at next-to-leading logarithmic (NLL$'$) accuracy that allows for the resummation of threshold logarithms and logarithms of the jet radius. See also~\cite{Procura:2011aq}, where the resummation was carried out at leading-logarithmic accuracy by making a suitable scale choice. We determine the relevant non-global logarithms (NGLs)~\cite{Dasgupta:2001sh} at next-to-next-to-leading order (NNLO) and resum them at leading-logarithmic accuracy in the leading-color approximation. The numerical size of the NGLs is significant near the hadronic threshold, and their resummation is formally necessary to achieve NLL$'$ accuracy. Overall we find that the threshold resummation has a significant numerical impact which motivates the extraction of threshold resummed fragmentation functions in the future.

We note that the longitudinal momentum distribution can be extended to the case where also the transverse momentum of hadrons relative to the jet axis is observed. In this case, the choice of the jet axis plays an important role. The transverse momentum measurement of hadrons relative to the standard jet axis was considered in~\cite{Bain:2016rrv,Kang:2017glf}, the Winnter-Take-All (WTA) scheme in~\cite{Neill:2016vbi,Neill:2018wtk}, and grooming was included in~\cite{Makris:2017arq}. Furthermore, spin asymmetries relevant for RHIC were considered in~\cite{Yuan:2007nd,Kang:2017btw,DAlesio:2017bvu}. An improved perturbative understanding of the longitudinal momentum distribution considered in this work will also advance these transverse momentum dependent jet substructure studies.

The remainder of this paper is organized as follows. In section~\ref{sec:NLO}, we review hadron-in-jet production in proton-proton collisions at NLO. We present the refactorization of the cross section at partonic threshold in section~\ref{sec:factorization} and list the results of the relevant functions at NLO. We solve the associated evolution equations in Mellin moment space which is presented in section~\ref{sec:resummation}. The final threshold and jet radius resummed cross section is given in section~\ref{sec:space} where we also include the contribution of NGLs. Numerical results are presented for LHC kinematics in section~\ref{sec:numerics} and we conclude in section~\ref{sec:conclusions} and present an outlook.

\section{Review of hadron-in-jet production in proton-proton collisions at NLO}
\label{sec:NLO}

We consider the longitudinal momentum distribution of hadrons inside jets in proton-proton collisions $pp\to({\rm jet}+h)+X$. The identified jets are reconstructed with the anti-k$_T$ algorithm~\cite{Cacciari:2008gp} differential in the jet transverse momentum $p_T$ and rapidity $\eta$. We denote the longitudinal momentum fraction of the identified hadrons inside the reconstructed jet by $z_h=p_T^h/p_T$, where $p_T^h$ denotes the hadron transverse momentum (relative to the beam axis). At NLO, up to power corrections ${\cal O}(R^2)$, we can write the cross section as~\cite{Kaufmann:2015hma,Kang:2016ehg}
\begin{equation}\label{eq:fact1}
\frac{{\rm d}\sigma^{pp\to({\rm jet}+h)X}}{{\rm d}p_T \,{\rm d}\eta\, {\rm d}z_h}=\sum_{abc}f_a(x_a,\mu)\,\otimes\, f_b(x_b,\mu)\,\otimes\, H_{ab}^c(x_a,x_b,\eta,p_T/z,\mu)\, \otimes\, {\cal G}_c^h(z,z_h,p_T R,\mu) \,.
\end{equation}
Here $f_{a,b}$ are the PDFs, $H_{ab}^c$ are the hard-scattering functions to produce a parton $c$ in the final state in a collision of partons $a$ and $b$. The hard functions are known analytically to NLO~\cite{Aversa:1988vb,Jager:2002xm}. The jet function ${\cal G}_c^h$ takes into account the formation and evolution of the observed final state jet which carries a momentum fraction $z=p_T/p_T^c$ of the initial parton $c$. In addition, the jet function contains the dependence on the variable $z_h$. Appropriate integrals over the longitudinal momentum fractions $x_{a,b}$ and $z$ are denoted by $\otimes$. In the kinematic regime away from the endpoints $0\ll z_h\ll 1$, we can write the jet function ${\cal G}_c^h$ in terms of a convolution integral of the standard collinear fragmentation functions $D_d^h$ and matching coefficients ${\cal J}_{cd}$ which can be calculated perturbatively,
\begin{equation}\label{eq:NLOfact}
{\cal G}_c^h(z,z_h,p_T R,\mu)=\sum_d{\cal J}_{cd}(z,z_h,p_T R,\mu)\otimes D_d^h(z_h,\mu)\,.
\end{equation}
The relevant expressions to NLO can be found in~\cite{Kaufmann:2015hma,Kang:2016ehg}. The factorization in Eqs.~(\ref{eq:fact1}) and~(\ref{eq:NLOfact}) is achieved by separating the physics at different scales, namely,
\begin{equation}
\mu_f\sim \Lambda_{\rm QCD}\,,\qquad \mu_{\cal G}\sim p_T R\,,\qquad \mu_H\sim p_T \,,
\end{equation}
of the PDFs and fragmentation function, the jet function and the hard function, respectively.  Single logarithms of the jet radius $\alpha_s^n\ln^n R^2$ can be resummed through DGLAP evolution equations satisfied by the jet function ${\cal G}_c^h$ which take the form
\begin{equation}\label{eq:dglap}
\mu\frac{{\rm d}}{{\rm d}\mu} {\cal G}_c^h(z,z_h,p_TR,\mu) = \sum_d \int_z^1\frac{{\rm d}z'}{z'} \frac{\alpha_s}{\pi} P_{dc}(z/z')\,{\cal G}_d^h(z',z_h,p_TR,\mu)\,.
\end{equation}
Here $P_{dc}$ are the usual Altarelli-Parisi splitting functions and the evolution is carried out between the scales $p_T R$ and $p_T$~\cite{Kang:2016ehg}. For completeness, we present here the results for the partonic unrenormalized jet function ${\cal G}_c^{q,g}$ and before subtracting the final state collinear singularity which is removed by introducing the nonperturbative fragmentation functions. We have~\cite{Kaufmann:2015hma,Kang:2016ehg}
\begin{subequations}\label{eq:GGrenqq}
\bea
\GG_{q}^q(z,z_h,p_T R,\mu)  = & \, \delta(1-z)\delta(1-z_h)
\nnu
& +\f{\as}{2\pi}\left(-\f{1}{\epsilon}-L\right) P_{qq}(z_h)\delta(1-z) +\f{\as}{2\pi} \left(\f{1}{\epsilon}+L\right)\, P_{qq}(z)\delta(1-z_h) 
\nnu
& + \delta(1-z)\f{\as}{2\pi}\left[2C_F(1+z_h^2)\left(\f{\ln(1-z_h)}{1-z_h}\right)_+ + C_F(1-z_h)+2P_{qq}(z_h)\ln z_h \right] \nnu
& - \delta(1-z_h)\f{\as}{2\pi}\left[2C_F(1+z^2)\left(\f{\ln(1-z)}{1-z}\right)_+ +C_F(1-z) \right] \, ,
\\
\GG_{q}^g(z,z_h,p_T R,\mu)  = & \, \f{\as}{2\pi}\left(-\f{1}{\epsilon}-L\right) P_{gq}(z_h)\delta(1-z) +\f{\as}{2\pi}\left(\f{1}{\epsilon}+L\right)\, P_{gq}(z)\delta(1-z_h) \nnu
& + \delta(1-z)\f{\as}{2\pi}\left[2P_{gq}(z_h)\ln(z_h(1-z_h))+ C_F z_h \right] \nnu
& - \delta(1-z_h)\f{\as}{2\pi}\left[2P_{gq}(z)\ln(1-z)+z \right] \, ,
\\
\GG_{g}^g(z,z_h,p_T R,\mu)  = & \, \delta(1-z)\delta(1-z_h)
\nnu
& +\f{\as}{2\pi}\left(-\f{1}{\epsilon}-L\right) P_{gg}(z_h)\delta(1-z) +\f{\as}{2\pi} \left(\f{1}{\epsilon}+L\right)\, P_{gg}(z)\delta(1-z_h) \nnu
& + \delta(1-z)\f{\as}{2\pi}\left[4C_A\f{(1-z_h+z_h^2)^2}{z_h}\left(\f{\ln(1-z_h)}{1-z_h}\right)_+ +2P_{gg}(z_h)\ln z_h \right] \nnu
& - \delta(1-z_h)\f{\as}{2\pi}\left[4C_A\f{(1-z+z^2)^2}{z}\left(\f{\ln(1-z)}{1-z}\right)_+ \right] \, ,
\\
\GG_{g}^q(z,z_h,p_T R,\mu)  = & \, \f{\as}{2\pi}\left(-\f{1}{\epsilon}-L\right) P_{qg}(z_h)\delta(1-z) +\f{\as}{2\pi}\left(\f{1}{\epsilon}+L\right)\, P_{qg}(z)\delta(1-z_h) \nnu
& + \delta(1-z)\f{\as}{2\pi}\left[2P_{qg}(z_h) \ln(z_h(1-z_h))+ 2T_F z_h(1-z_h) \right] \nnu
& - \delta(1-z_h)\f{\as}{2\pi}\left[2P_{qg}(z)\ln(1-z)+2T_F z(1-z) \right] \, .
\eea
\end{subequations}
where
\be
L=\ln\left(\f{\mu^2}{p_T^2 R^2}\right) \,.
\ee
Note that the matching coefficients ${\cal J}_{cd}$ at NLO in Eq.~(\ref{eq:NLOfact}) are obtained from Eq.~(\ref{eq:GGrenqq}) by removing all $1/\epsilon$ singularities. We also note that it is possible to further separate the NLO result into two jet functions where each function only depends on $z$ and $z_h$, respectively. See~\cite{Kaufmann:2015hma} for more details. Note that each jet function ${\cal G}_c^{q,g}$ contains two logarithms of the jet radius $\sim L$. The logarithms multiplied by $\sim P_{dc}(z)$ are resummed through the DGLAP evolution of the jet function in Eq.~(\ref{eq:GGrenqq}). The other ones which are multiplied by $\sim P_{dc}(z_h)$ are resummed by evaluating the fragmentation function at the jet scale $\mu_J\sim p_T R$, see~\cite{Kang:2016ehg}, whereas at fixed order the scale is chosen as $\mu_H\sim p_T$. Thus, the two DGLAP evolutions achieve the resummation of both sets of logarithms of the jet radius $R$. The relevant results for the cone algorithm can be found in~\cite{Kang:2017mda}. We note that the threshold resummation discussed in this work does not depend on the choice of the jet algorithm to the accuracy we are working at.

\section{Refactorization at partonic threshold} 
\label{sec:factorization}

We consider the threshold resummation for hadron-in-jet production in the limit $z_h\to 1$. This corresponds to the limit where the identified hadron inside the jet carries most of the jet transverse momentum and additional emissions inside the jet are soft. Note that we do not take the limit $z\to 1$ which is associated with the threshold production of the jet, see for example~\cite{Kidonakis:2000gi,deFlorian:2007fv,Becher:2009th,deFlorian:2013qia,Catani:2013vaa,Dai:2017dpc,Liu:2017pbb,Hinderer:2018nkb}. The refactorization at threshold is obtained within Soft Collinear Effective Theory (SCET)~\cite{Bauer:2000ew,Bauer:2000yr,Bauer:2001ct,Bauer:2001yt,Beneke:2002ph}. We start with the fixed order result of the jet function ${\cal G}_c^d$ in the threshold limit. By taking the limit $z_h\to 1$, we find at NLO
\begin{subequations}\label{eq:Gthreshold}
\bea
\GG_{q}^q(z,z_h,p_T R,\mu)  = & \, \delta(1-z)\delta(1-z_h)\nnu
& +\f{\as}{2\pi}\left(-\f{1}{\epsilon}-L\right) C_F \left(\f{2}{(1-z_h)}_++\f32 \delta(1-z_h)\right)\delta(1-z)\nnu
& +\f{\as}{2\pi} \left(\f{1}{\epsilon}+L\right)\, P_{qq}(z)\delta(1-z_h) + \delta(1-z)\f{\as}{2\pi} 4C_F\left(\f{\ln(1-z_h)}{1-z_h}\right)_+   \nnu
& - \delta(1-z_h)\f{\as}{2\pi}\left[2C_F(1+z^2)\left(\f{\ln(1-z)}{1-z}\right)_+ +C_F(1-z)\right] \, ,
\\
\GG_{q}^g(z,z_h,p_T R,\mu)  = & \, \delta(1-z_h)\f{\as}{2\pi}\left[\left(\f{1}{\epsilon}+L\right) P_{gq}(z) - 2P_{gq}(z)\ln(1-z)-C_F z \right] \,,
\\
\GG_{g}^g(z,z_h,p_T R,\mu)  = & \, \delta(1-z)\delta(1-z_h)\nnu
& +\f{\as}{2\pi}\left(-\f{1}{\epsilon}-L\right) C_A\left(\frac{2}{(1-z_h)}_++\frac{\beta_0}{2}\delta(1-z_h)\right)\delta(1-z)  \nnu
& +\f{\as}{2\pi}\left(\f{1}{\epsilon}+L\right)\, P_{qg}(z)\delta(1-z_h) + \delta(1-z)\f{\as}{2\pi} 4C_A\left(\f{\ln(1-z_h)}{1-z_h}\right)_+ \nnu
& - \delta(1-z_h)\f{\as}{2\pi}\left[4C_A\f{(1-z+z^2)^2}{z}\left(\f{\ln(1-z)}{1-z}\right)_+ \right] \, ,
\\
\GG_{g}^q(z,z_h,p_T R,\mu)  = & \, \delta(1-z_h)\f{\as}{2\pi}\left[\left(\f{1}{\epsilon}+L\right) P_{qg}(z)-2P_{qg}(z)\ln(1-z)-2T_F z(1-z) \right] \,.
\eea
\end{subequations}
Note that the jet function in the threshold limit $z_h\to 1$ still satisfies the DGLAP evolution equations in Eq.~(\ref{eq:dglap}) as it is associated with the variable $z$. In the threshold limit $\Lambda_{\rm QCD}\ll p_TR(1-z_h)\ll p_TR$ at NLL$'$ accuracy, we can refactorize the jet function ${\cal G}_c^h$ in Eq.~(\ref{eq:fact1}) as
\be\label{eq:refactorize}
{\cal G}_c^h(z,z_h,p_T R,\mu) \stackrel{{\rm NLL}'}{=}\sum_d {\cal H}_{cd}(z,p_T R,\mu)\, S_d(z_h,p_TR,\mu)\otimes S_d^{\rm NG}(z_h,\mu)\otimes D_d^h(z_h,\mu)\,.
\ee
Here ${\cal H}_{cd}$ are hard matching coefficients functions. To NLO they are given by out-of-jet radiation diagrams and the characteristic scale is given by the jet scale $\mu_{\cal H}\sim \mu_{\cal G}\sim p_T R$. In the limit $z_h\to 1$ any emissions inside the observed jet are required to be soft. Soft radiation inside the jet is taken into account by a collinear-soft~\cite{Bauer:2011uc} function $S_d$ where the collinearity is set by the jet constraint. The characteristic soft scale $\mu_S$ is given in terms of the threshold variable $\mu_S\sim p_T R(1-z_h)$. The evolution equation of the soft function which we derive below therefore resums the relevant threshold logarithms. We account for NGLs~\cite{Dasgupta:2001sh} by including the soft function $S_{d}^{\rm NG}$ which can be separated from the soft function $S_{d}$ at NLL$'$ accuracy. NGLs arises due to correlations between the out-of-jet radiation ${\cal H}_{cd}$ and the soft function $S_d$. They first arise at NNLO and need to be taken into account to achieve full NLL$'$ accuracy. See for example~\cite{Hatta:2013iba,Becher:2015hka,Neill:2016stq,Larkoski:2016zzc,Balsiger:2019tne} for recent work on the all order resummation of NGLs. We take into account the parton-to-hadron fragmentation by introducing the fragmentation function $D_d^h$ at threshold. The symbols $\otimes$ denote Mellin convolution integrals in terms of $z_h$ similar to those in Eq.~(\ref{eq:NLOfact}) above.

Next we calculate the relevant functions in the refactorized expression in Eq.~(\ref{eq:refactorize}) to NLO. Throughout this work, we use the $\overline{\rm MS}$ scheme. An important check is that the one-loop results add up to the fixed order result in Eq.~(\ref{eq:Gthreshold}). We start with the hard matching coefficients functions ${\cal H}_{cd}$ at the jet scale. The same functions appear in the refactorization of the cross section of other jet substructure observables which were first calculated in~\cite{Kang:2017glf,Kang:2017mda}. For completeness, we list the results here including the singularities:
\begin{subequations}
\begin{align}\label{eq:Hcd}
 {\cal H}_{qq}(z,p_{T} R,\mu) 
  &= \delta(1-z) + \f{\as}{2\pi} \bigg[C_F \delta(1-z)\Big(-\f{1}{\epsilon^2}-\f{3}{2\epsilon}-\frac{L}{\epsilon}-\f{L^2}{2}  - \frac32 L +\f{\pi^2}{12} \Big) 
\nnu
 & \quad
+\Big(\f{1}{\epsilon}+L\Big)\, P_{qq}(z) -2C_F(1+z^2)\Big(\f{\ln(1-z)}{1-z}\Big)_+ -C_F(1-z)  \bigg] 
\,, \\
 {\cal H}_{qg}(z,p_{T} R,\mu) 
 &=\f{\as}{2\pi}\bigg[\Big(\frac{1}{\epsilon}+L\Big)P_{gq}(z) - 2 P_{gq}(z)\ln(1-z) - C_Fz \bigg]
\,, \\
{\cal H}_{gg}(z, p_{T} R, \mu) 
& = \delta(1-z) + \f{\as}{2\pi}\bigg[C_A \delta(1-z)\Big(-\frac{1}{\epsilon^2}-\f{\beta_0}{2C_A\epsilon}-\frac{1}{\epsilon}L-\f{L^2}{2} - \f{\beta_0}{2C_A} L + \frac{\pi^2}{12}\Big)
\nnu 
& \quad
+ \Big(\f{1}{\epsilon}+L\Big)\, P_{gg}(z) - \frac{4C_A (1-z+z^2)^2}{z} \left(\frac{\ln(1-z)}{1-z}\right)_{+} \bigg]
\,,\\
{\cal H}_{gq}(z,p_{T} R, \mu) 
 & =  \f{\as}{2\pi}\bigg[\Big(\frac{1}{\epsilon}+L\Big)P_{qg}(z) - 2 P_{qg}(z)\ln(1-z) - T_F 2z(1-z) \bigg]\,.
\end{align}
\end{subequations}
The perturbative parton-to-parton fragmentation functions at threshold are given by~\cite{Stewart:2010qs}
\ba\label{eq:FFNLO}
D_q(z_h,\mu) &=& \delta(1-z_h)-\f{\as}{2\pi}C_F \f{1}{\epsilon} \left(\f{2}{(1-z_h)}_++\f32 \delta(1-z_h)\right)\,,\\
D_g(z_h,\mu) &=& \delta(1-z_h)-\f{\as}{2\pi}C_A \f{1}{\epsilon} \left(\f{2}{(1-z_h)}_++\f{\beta_0}{2C_A} \delta(1-z_h)\right)\,,
\ea
for quarks and gluons. Note that the singularities here match the ones in Eq.~(\ref{eq:Gthreshold}). We define the collinear-soft function for quarks at the operator level as
\be
S_q(q^-,p_T R,\mu) = \frac{1}{N_c}\sum_{X_s} \langle 0  |  {\rm\bar T}[Y_{\bar n}^\dagger Y_n]\, |X\rangle \langle X | {\rm T}[Y_n^\dagger Y_{\bar n} ]  |  0  \rangle\, \delta\Big(q^- - \sum_{i\in {\rm jet}} q_i^-\Big)\,,
\ee
where the delta function sums over the light-cone component $q^-$ of the soft radiation inside the jet. Soft emissions from collinear particles in the jet are taken into account by the eikonal Wilson line $Y_n$ and we include a Wilson line in the conjugate direction $\bar n$. For gluons, we need the Wilson lines in the adjoint representation and the overall normalization needs to be adjusted to $1/(N_c^2-1)$. The NLO collinear-soft function can be obtained from
\be
S_d^{(1)}(q^-,p_TR,\mu)= C_d \f{g_s^2}{(2\pi)^{d-1}}\left(\f{\mu^2 e^{\gamma_E}}{4\pi}\right)^\epsilon \int {\rm d}^dk\, \delta(k^2) \f{n\cdot\bar n}{n\cdot k\, \bar n\cdot k}\delta(q^- -\, \bar n\cdot k)\,\Theta\!\left(\f{n\cdot k}{\bar n\cdot k}<\f{R^2}{4}\right) \,.
\ee
for quarks and gluons $C_d=C_{F,A}$. The NLO calculation turns out to be analogous to that for the soft-collinear function considered in~\cite{Dai:2017dpc,Liu:2017pbb}. The ${\cal O}(\alpha_s)$ contribution of the two functions only differs by an overall minus sign. We find
\be
S_d^{(1)}(q^-,p_TR,\mu)= \f{\as}{\pi} C_d \left(-\f{1}{\epsilon}\right) \f{e^{\epsilon\gamma_E}}{\Gamma[1-\epsilon]}\f{R}{\mu}\left(\f{q^- R}{\mu}\right)^{-1-2\epsilon} \,.
\ee
At threshold we can identify $q^-\approx (1-z_h)p_T$. We therefore write the soft function as
\be
S_d^{(1)}(z_h,p_TR,\mu)= \f{\as}{\pi} C_d \left(-\f{1}{\epsilon}\right) \f{e^{\epsilon\gamma_E}}{\Gamma[1-\epsilon]}\left(\f{\mu}{p_T R}\right)^{2\epsilon} \left(1-z_h \right)^{-1-2\epsilon} \,.
\ee
The expansion of the soft function up to NLO in distribution space is then given by
\begin{align}\label{eq:softNLO}
S_d(z_h,p_T R,\mu) & =\delta(1-z_h)+\f{\as}{2\pi} C_d \,\delta(1-z_h)\left[\f{1}{\epsilon^2}+\f{L}{\epsilon}+\f{L^2}{2}-\f{\pi^2}{12}\right]
\nnu
& \quad +\f{\as}{2\pi} C_d \left[-\Big(\f{1}{\epsilon}+L\Big)\f{2}{(1-z_h)}_+ + 4 \Big(\f{\ln(1-z_h)}{1-z_h}\Big)_+\right] \,.
\end{align}
We observe that the ${\cal O}(\alpha_s)$ contributions of the different functions in Eq.~(\ref{eq:Hcd}),~(\ref{eq:FFNLO}) and~(\ref{eq:softNLO}) add up to the fixed order result in the threshold limit as given in Eq.~(\ref{eq:Gthreshold}), which provides an important check of the refactorization. Up to power corrections of order ${\cal O}(R^2)$, the leading non-global logarithms for hadron-in-jet production are a boosted version of the $e^+e^-$ hemisphere case~\cite{Banfi:2010pa}. We find that the leading NGLs at NNLO are given by
\begin{equation}
S^{\rm NG}_d(z_h,\mu)=\delta(1-z_h)-\frac{\alpha_s^2 C_AC_d}{6} \Big(\frac{\ln(1-z_h)}{1-z_h}\Big)_+ \,.
\end{equation}
In our numerical results presented below, we include the all order resummation of the NGLs at leading logarithmic accuracy in the leading-color approximation~\cite{Dasgupta:2001sh}.

\section{Evolution in Mellin space}
\label{sec:resummation}

We start with the evolution equations of the hard-matching functions ${\cal H}_{cd}$ which are given by~\cite{Kang:2017glf,Kang:2017mda}
\begin{equation}
\mu\frac{{\rm d}}{{\rm d}\mu}{\cal H}_{cd}(z,p_TR,\mu)=\sum_e\int_z^1\frac{{\rm d}z'}{z'}\gamma_{ce}^{\cal H}\Big(\frac{z'}{z},p_T R,\mu\Big)\,{\cal H}_{ed}(z',p_T R,\mu)\,.
\end{equation}
The anomalous dimensions $\gamma_{ce}^{\cal H}$ are given by
\begin{equation}
\gamma_{ce}^{\cal H}(z,p_TR,\mu)=\delta_{ce}\delta(1-z)\Gamma^{\cal H}_c(p_T R,\mu)+\frac{\alpha_s}{\pi}P_{ec}(z) \,,
\end{equation}
where $P_{ec}$ are the Altarelli-Parisi splitting functions and the additional diagonal terms are
\begin{align}\label{eq:Hijdiag}
\Gamma^{\cal H}_q(p_T R,\mu) & = \frac{\alpha_s}{\pi} C_F \bigg(-L-\frac32 \bigg)\,, \\
\Gamma^{\cal H}_g(p_T R,\mu) & = \frac{\alpha_s}{\pi} C_A \bigg(-L-\frac{\beta_0}{2C_A} \bigg) \,.
\end{align}
We are going to derive the evolution equations of the soft function and the fragmentation functions in Mellin moment space where the convolution structure in Eq.~(\ref{eq:refactorize}) factorizes. After solving the evolution equations and performing the resummation in Mellin transform space, we will perform a numerical Mellin inverse transformation. We adopt the following convention for the Mellin transform and its inverse
\begin{align}
f(N) & =\int_0^1 {\rm d}z \, z^{N-1} f(z) \,, \\
f(z)  & = \int_{{\cal C}_N}\f{{\rm d}N}{2\pi i} z^{-N}f(N)\,,
\end{align}
where ${{\cal C}_N}$ denotes an appropriate contour in the complex $N$ plane. For the soft function in Mellin space we find
\begin{align}
S_d(p_T R/\bar N,\mu) & =1+\f{\as}{2\pi} C_d \left[\f{1}{\epsilon^2}+\f{2}{\epsilon}\ln\left(\f{\mu\bar N}{p_T R}\right)+2 \ln^2\left(\f{\mu\bar N}{p_T R}\right)+\f{\pi^2}{4}\right]\,,
\end{align}
where we use the notation $\bar N=Ne^{\gamma_E}$. For the fragmentation function at threshold we obtain in Mellin space
\ba
D_q(\bar N,\mu) &=& 1+\f{\as}{2\pi}C_F \f{1}{\epsilon} \left(2\ln\bar N -\f32\right)\,,\\
D_g(\bar N,\mu) &=& 1+\f{\as}{2\pi}C_A \f{1}{\epsilon} \left(2\ln\bar N -\f{\beta_0}{2C_A}\right)\,.
\ea
The characteristic scales of the fragmentation function $D_d$, the soft function $S_d$ and the hard matching functions ${\cal H}_{cd}$ are thus given by
\be
 \mu_D\sim\Lambda_{\rm QCD}\,\qquad \mu_S\sim p_T R/\bar N \,,\qquad \mu_{\cal H}\sim\mu_{\cal G}\sim p_T R \,.
\ee
In Mellin moment space the renormalization and the evolution equations of the soft function are multiplicative. We can relate the bare and renomalized soft functions as
\be
S_d(p_T R/\bar N,\mu)=Z_d(p_T R/\bar N,\mu)\, S_d^{\rm ren} (p_T R/\bar N,\mu) \,.
\ee
From now on we only consider the renormalized soft function and drop its superscript ``ren''. The renormalization constant $Z_d$ and the renormalized soft function $S_d$ are given by
\begin{align}
Z_d(p_T R/\bar N,\mu) & = 1+\f{\as}{2\pi} C_d \left[\f{1}{\epsilon^2}+\f{2}{\epsilon}\ln\left(\f{\mu\bar N}{p_T R}\right)\right] \,, \\
S_d(p_T R/\bar N,\mu)  & = 1+\f{\as}{2\pi} C_d \left[2 \ln^2\left(\f{\mu\bar N}{p_T R}\right)+\f{\pi^2}{4}\right]\,.\label{eq:Sqren}
\end{align}
The renormalized soft function satisfies the following renormalization group (RG) evolution equation
\be\label{eq:SqRG}
\mu\f{{\rm d}}{{\rm d}\mu}S_d(p_T R/\bar N,\mu) =\gamma_d^S(p_T R/\bar N,\mu) \, S_d(p_T R/\bar N,\mu) \,,
\ee
where the anomalous dimension is given by
\be
\gamma_d^S(p_T R/\bar N,\mu)=2 \f{\as}{\pi} C_d \ln\left(\f{\mu\bar N}{p_T R}\right) \,.
\ee
To achieve NLL$'$ accuracy, we need to include the two loop contribution of the cusp anomalous dimension. We thus write $\gamma_d^S$ as
\be
\gamma_d^S(p_T R/\bar N,\mu)=2 \Gamma_d(\alpha_s) \ln\left(\f{\mu\bar N}{p_T R}\right) + \gamma_{d}(\alpha_s)\,.
\ee
Here $\gamma_d$ is the non-cusp term which is zero at NLL$'$ accuracy and $\Gamma_d(\alpha_s)$ is the cusp anomalous dimension which can be calculated perturbatively as
\be\label{eq:cusp}
\Gamma_d(\alpha_s) = \frac{\alpha_s}{\pi}\Gamma_d^{(1)} + \left(\frac{\alpha_s}{\pi}\right)^2 \Gamma_d^{(2)} + \ldots \,.
\ee
To achieve NLL$'$ accuracy, we need the first two coefficients which are given by
\be
\Gamma_d^{(1)}=C_d\,,\qquad \Gamma_d^{(2)}=\frac{C_d}{4}\left[\left(\frac{67}{9}-\frac{\pi^2}{3}\right)C_A-\frac{20}{9} T_F n_f \right] \,,
\ee
with $C_d=C_{F,A}$ for quarks and gluons, respectively. The solution of the RG equation of the soft function in Eq.~(\ref{eq:SqRG}) can be written as
\be\label{eq:softexp}
S_d(p_T R/\bar N,\mu)=S_d(p_T R/\bar N,\mu_0)\exp\left[\int_{\mu_0}^\mu {\rm d}\!\ln\mu' \,\gamma_d^S(p_T R/\bar N,\mu') \right] \,.
\ee
We note that at leading-logarithmic accuracy for a fixed coupling constant, the exponential in Eq.~(\ref{eq:softexp}) can be written as
\begin{equation}\label{eq:LL}
\exp\left[\frac{\alpha_s}{\pi}C_d \ln^2\bar N \right] \,.
\end{equation}
Next, we consider the RG evolution equations of the fragmentation functions at threshold. They satisfy the usual DGLAP evolution equations in the threshold limit. In Mellin space, we have
\be\label{eq:evolFF}
\mu\f{{\rm d}}{{\rm d}\mu} D_d(\bar N,\mu) = \gamma_d^D(\bar N,\mu)\, D_d(\bar N,\mu)\,,
\ee
where the anomalous dimensions are given by~\cite{Stewart:2010qs}
\ba\label{eq:dimFF}
\gamma_q^D(\bar N,\mu) &=& \f{\as}{\pi}C_F\left[-2\ln\bar N+\f32 \right]\,, \\
\gamma_g^D(\bar N,\mu) &=& \f{\as}{\pi}C_A\left[-2\ln\bar N+\frac{\beta_0}{2C_A} \right] \,.
\ea
The solution of the RG evolution equation of the fragmentation function in Eq.~(\ref{eq:evolFF}) can be written analogously to that for the resummed soft function in Eq.~(\ref{eq:softexp}). We find that the RG invariance of the refactorized cross section in the threshold limit is satisfied,
\be\label{eq:RGconsistency}
\Gamma_d^{{\cal H}}(p_T R,\mu)+\gamma_d^{S}(p_T R/\bar N,\mu)+\gamma_d^{D}(\bar N,\mu)=0 \,,
\ee
where we included $\Gamma_d^{{\cal H}}$ which denotes the diagonal non-DGLAP part of the anomalous dimensions of the hard matching functions as given in Eq.~(\ref{eq:Hijdiag}). The RG invariance in Eq.~(\ref{eq:RGconsistency}) implies that indeed between the scale $p_T R$ and $p_T$ only the DGLAP evolution of the entire jet function ${\cal G}_c$ remains, see Eq.~(\ref{eq:dglap}). From Eq.~(\ref{eq:Sqren}), we find that choosing the natural scale of the soft function $\mu_S\sim p_T R/\bar N$ eliminates all large logarithms at fixed order. In order to perform the Mellin inverse transformation, we need to choose a prescription to avoid the Landau pole as discussed in the next section.

\section{Threshold and jet radius joint resummation at NLL$'$}
\label{sec:space}

When performing the Mellin inverse transformation we need to avoid the Landau pole. This can be achieved for example by  using the minimal prescription of~\cite{Catani:1996yz} where we choose the scale $\mu_0=\mu_S\sim p_T R/\bar N$ in Mellin space. The contour in the complex plane of the Mellin inverse transformation is chosen to the right of all the singularities of the resummed cross section and to the left of the branch cut due to the Landau pole. Alternatively, we could follow for example the prescription developed in~\cite{Becher:2006nr} where the scale setting is carried out in distribution space. We can evaluate the integral in Eq.~(\ref{eq:softexp}) by inserting an explicit form for the QCD strong coupling constant $\alpha_s$. We use
\begin{equation}
\alpha_s(\mu)=\frac{\alpha_s(\mu')}{X}\left[1-\alpha_s(\mu')\frac{b_1}{b_0}\frac{\ln X}{X}\right] \,,
\end{equation}
with
\begin{equation}
X=1+b_0\alpha_s(\mu')\ln\Big(\frac{\mu^2}{\mu'^2}\Big) \,,
\end{equation}
which is sufficient to NLL$'$ accuracy. See also~\cite{Czakon:2018nun} for example. The first two coefficients of the QCD beta function are given by
\begin{equation}
b_0=\frac{1}{12\pi}(11 C_A-2N_f),\,\qquad b_1= \frac{1}{24\pi^2}(17C_A^2-5C_AN_f-3C_FN_f)\,.
\end{equation}
In order to achieve the resummation we treat $\alpha_s\ln\bar N\sim {\cal O}(1)$ and expand the remainder in powers of $\alpha_s$. We find
\begin{equation}\label{eq:softexponential}
S_d(p_T R/\bar N,\mu)=S_d(p_T R/\bar N,\mu_S)\exp\left[\frac{\lambda}{\alpha_s b_0}\,h_d^{(1)}(\lambda)+h_2^{(2)}(\lambda,p_TR,\mu)\right] \,,
\end{equation}
where 
\begin{equation}
\lambda = \alpha_s b_0\ln \Big(\frac{\mu_{\cal H}}{\mu_S}\Big)=\alpha_s b_0\ln\bar N \,,
\end{equation}
and
\begin{eqnarray}
h_d^{(1)}(\lambda)&=&\frac{\Gamma_d^{(1)}}{2\pi b_0\lambda}[2\lambda+(1-2\lambda)\ln(1-2\lambda)] \,,\\
h_d^{(2)}(\lambda,p_T R,\mu)&=&-\frac{\Gamma_d^{(2)}}{2\pi^2b_0^2}[2\lambda+\ln(1-2\lambda)]\nn \\
&+&\frac{\Gamma_d^{(1)}b_1}{2\pi b_0^3}\Big[2\lambda+\ln(1-2\lambda)+\frac12\ln^2(1-2\lambda)\Big]\nn\\
&-&\frac{\Gamma_d^{(1)}}{2\pi b_0}\ln(1-2\lambda)\ln \Big(\frac{\mu^2}{p_T^2R^2}\Big) \,.
\end{eqnarray}
The relevant scale here is $\mu\sim\mu_{\cal H}\sim\mu_{\cal G}\sim p_T R$ as we need to evolve the soft function to the jet scale. Above the jet scale up to the hard scale $\mu_H\sim p_T$ we are only left with the DGLAP evolution equations of the entire jet function in Eq.~(\ref{eq:dglap}) which resums logarithms of the jet radius, see Eq.~(\ref{eq:RGconsistency}). Therefore, the result in Eq.~(\ref{eq:softexponential}) only exponentiates threshold logarithms but not logarithms of the jet radius $R$. Instead, logarithms of the jet radius are resummed through the DGLAP evolution of the jet function ${\cal G}_c^h$ in Eq.~(\ref{eq:dglap}) and the DGLAP evolution of the fragmentation function up the scale $p_T R$. In order to compare to the fixed order expression, we need to evaluate the fragmentation functions instead at the hard scale $\mu_H\sim p_T$. We can thus subtract the perturbative evolution of the fragmentation function in the threshold limit between $p_T R$ and $p_T$. In the threshold limit, we would then include the factor
\be\label{eq:thresholdlnR}
\exp\bigg[\frac{\Gamma_d^{(1)}}{\pi b_0}\lambda\ln \Big(\frac{1}{R^2}\Big) \bigg]\,,
\ee
in the resummed result which can be derived from Eqs.~(\ref{eq:evolFF}) and~(\ref{eq:dimFF}). Note that using Eq.~(\ref{eq:thresholdlnR}) does not jointly resum threshold and jet radius logarithms but instead only logarithms of $R$ which are multiplied by a threshold logarithm. Nevertheless, Eq.~(\ref{eq:thresholdlnR}) provides an important check that indeed all threshold distributions are resummed within the our formalism. Expanding the exponential in Eq.~(\ref{eq:thresholdlnR}) to ${\cal O}(\alpha_s)$ correctly reproduces the terms $\sim -L P_{de}(z_h)$ of the NLO result at threshold, see Eq.~(\ref{eq:Gthreshold}). Finally, note that the remaining soft function evaluated at the characteristic scale still contains the strong coupling constant evaluated at the soft scale $\alpha_s(\mu_S\sim p_TR/\bar N)$. At NLL$'$ accuracy, we can rewrite $\alpha_s$ as~\cite{Hinderer:2014qta}
\be
\alpha_s(p_TR/\bar N)=\frac{\alpha_s(p_T R)}{1-2\lambda} \,.
\ee
We note that the threshold resummed exponent in Eq.~(\ref{eq:softexponential}) has the form as it is typically found in the ``direct'' perturbative QCD literature~\cite{Sterman:1986aj,Catani:1989ne}. We can also rewrite Eq.~(\ref{eq:softexp}) in the more conventional form of~\cite{Manohar:2003vb}
\begin{equation}
S_d(p_T R/\bar N,\mu)=S_d(p_T R/\bar N,\mu_S)\exp\left[\int_0^1{\rm d}z\frac{z^{N}-1}{1-z}\int_{\mu^2}^{p_T^2 R^2(1-z)^2}\frac{{\rm d}k^2}{k^2} \Gamma_d(\alpha_s(k^2))\right] \,,
\end{equation}
\begin{figure}[t]
\centering
\includegraphics[width=0.7\textwidth]{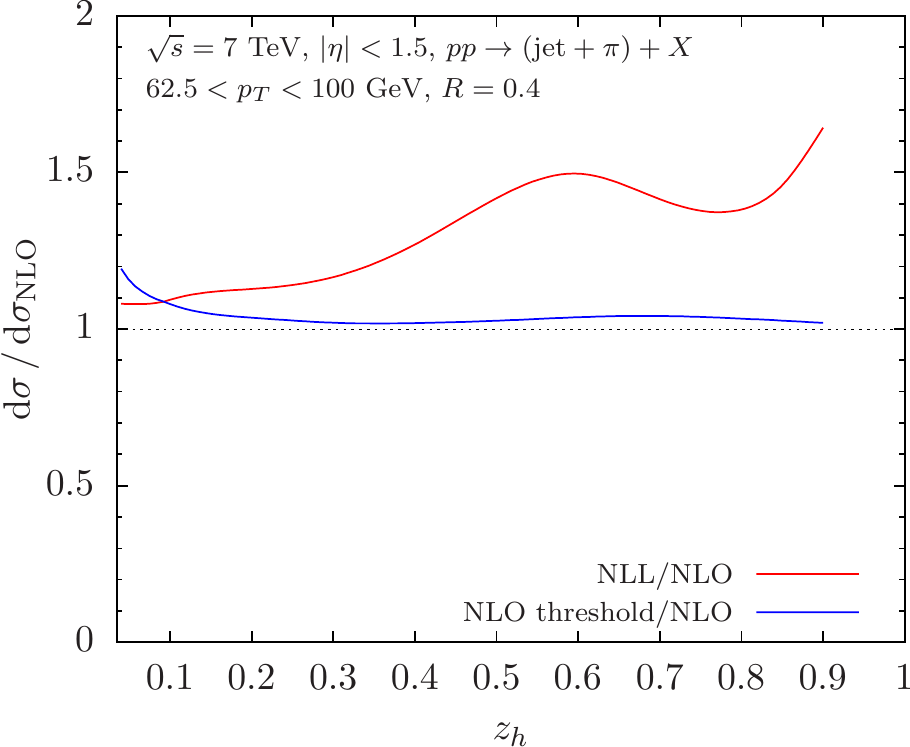}  
\caption{Ratio of the threshold resummed cross section and the NLO one (red) for the longitudinal momentum distribution $z_h$ of pions inside a jet $pp\to({\rm jet}+\pi)+X$. The kinematics are indicated in the figure and we use the pion fragmentation functions of~\cite{deFlorian:2014xna}. In addition, we show in blue the ratio of the NLO cross section in the threshold limit and the full NLO.~\label{fig:NLORESTHR}} 
\end{figure}
where we omitted the non-cusp contribution which only contributes beyond NLL$'$ accuracy. We include the non-global logarithms in Mellin space using the MC results of~\cite{Dasgupta:2001sh}. We introduce the variable
\begin{equation}
t=\frac{1}{2\pi}\int _{p_T R/\bar N}^{p_T R}\frac{{\rm d}k_T}{k_T}\alpha_s(k_T)=-\frac{\ln(1-2\lambda)}{4\pi b_0} \,,
\end{equation}
where $\lambda$ is defined as above. We can then write the non-global component at leading logarithmic accuracy in the large-$N_c$ limit as~\cite{Dasgupta:2001sh}
\begin{equation}
S_d^{\rm NG}(N,\mu)=\exp\left[-C_dC_A\frac{\pi^2}{3}\frac{1+(at)^2}{1+(bt)^c} t^2\right] \,,
\end{equation}
with $a=0.85\, C_A$, $b=0.86\, C_A$ and $c=1.33$. Eventually, we obtain the resummed result in $z_h$-space as
\begin{equation}\label{eq:Minverse}
{\cal G}_c^h(z,z_h,p_T R,\mu) = \sum_d {\cal H}_{cd}(z,p_T R,\mu)\int_{{\cal C}_N} \frac{{\rm d}N}{2\pi i}z_h^{-N} S_d(N,p_TR,\mu)\, S_d^{\rm NG}(N,\mu)\, D_d^h(N,\mu) \,.
\end{equation}
The involved fragmentation functions are fitted using a functional form suitable for the Mellin transformation as discussed in more detail below. We use an additive matching procedure to NLO,
\begin{equation}
{\rm d}\sigma^{\rm res}+{\rm d}\sigma^{\rm NLO}-{\rm d}\sigma^{\rm NLO\; threshold}\,,
\end{equation}
where we subtract the NLO contribution in the threshold limit or equivalently the ${\cal O}(\alpha_s)$ expansion of the resummed result to avoid double counting.

\section{Numerical results}
\label{sec:numerics}

\begin{figure}[t]
\centering
\includegraphics[width=0.7\textwidth]{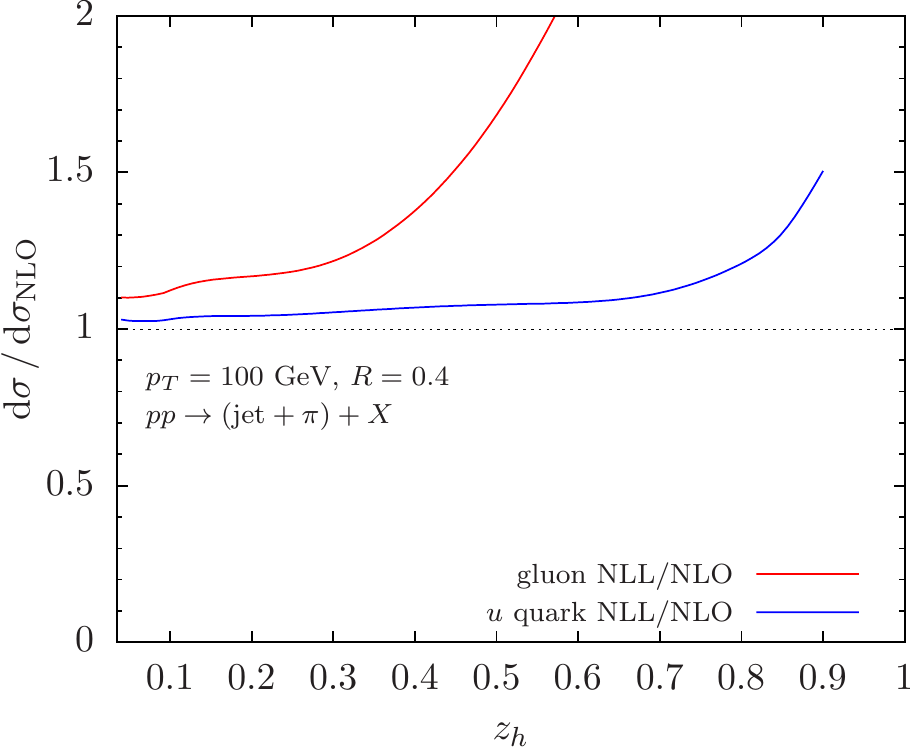}  
\caption{Ratio of the threshold resummed cross section and the NLO one for the gluon (red) and the $u$ quark jet function (blue). The kinematics are the same as in Fig.~\ref{fig:NLORESTHR}.~\label{fig:RESQG}} 
\end{figure}

We perform the Mellin inverse transformation numerically according to Eq.~(\ref{eq:Minverse}). Since fragmentation functions are usually provided on a grid in terms of the momentum fraction $z_h$ and the scale $\mu$, we first fit the fragmentation functions using a suitable functional form. We choose
\begin{equation}\label{eq:fitFF}
D^h_d(z_h,\mu)=a_1 z_h^{a_2}(1-z_h)^{a_2}\bigg( \sum_{j=0}^4 c_j\, z_h^j \bigg)\,,
\end{equation}
where the $a_i$ are given by
\begin{equation}
a_i=a_{i1}+a_{i2}\ln(\ln(\mu/\mu_0))
\end{equation}
and similarly for $c_j$. Here $\mu_0$ is a fixed parameter of order 1~GeV and the remaining parameters are fitted to the fragmentation functions. We can then calculate the Mellin transform of the fitted fragmentation functions analytically. In Mellin space the fitted fragmentation functions in Eq.~(\ref{eq:fitFF}) are given by a sum over Beta functions. In this work, we choose the DSS14/17 set of pion fragmentation functions (charge averaged) of~\cite{deFlorian:2014xna,deFlorian:2017lwf} as an example. We assess the quality of our fit in Eq.~(\ref{eq:fitFF}) by comparing the cross section at NLO using the fitted fragmentation functions in Mellin space and the exact ones in $z_h$-space. We find that the two results agree to better than 2\% in the relevant kinematic range considered below. Since the fragmentation functions that are currently available in the literature are extracted within a global analysis at fixed order in perturbation theory, we do not necessarily expect to find a better agreement with the data from the LHC and RHIC. Therefore, we limit ourselves here to studies of the numerical impact of the threshold resummation. However, in the future it will be possible to include the resummation within a global analysis and obtain threshold resummed fragmentation functions. 

\begin{figure}[t]
\centering
\includegraphics[width=0.7\textwidth]{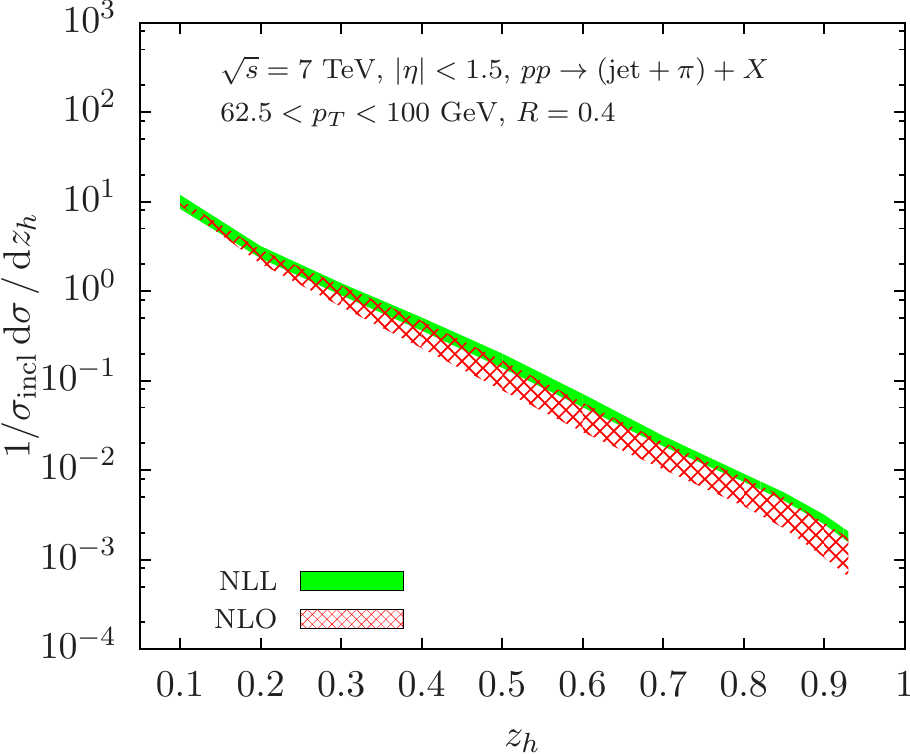}  
\caption{QCD scale uncertainty of the threshold resummed cross section (green) and the NLO result (red) normalized by the inclusive jet cross section $\sigma_{\rm incl}$.~\label{fig:NLORESscale}} 
\end{figure}

We consider inclusive jet production where the longitudinal momentum distribution of charged pions (but averaged over charge) is measured inside the observed jet, $pp\to({\rm jet}+\pi)+X$. As an example, we choose jets that are reconstructed with the anti-k$_T$ algorithm with a jet radius of $R=0.4$ in the transverse momentum range $62.5<p_T<100$~GeV with $|\eta|<1.5$ at $\sqrt{s}=7$~TeV. In Fig.~\ref{fig:NLORESTHR}, we show the ratio of the threshold resummed cross section and the result at NLO (red). Note that $\ln R$ resummation is included both in the numerator and denominator. The numerical impact of $\ln R$ resummation was studied in~\cite{Kang:2016ehg}. In addition, we show the ratio of the NLO result in the threshold approximation and the full NLO (blue). In the displayed range, the two calculations agree to better than 20\% and above $z_h=0.1$ to much better than 10\% indicating that the cross section is dominated by the singular distributions at partonic threshold. We observe that the threshold resummation leads to a sizable increase of the cross section in the intermediate and large $z_h$ region. Interestingly, we observe a peak at intermediate values of the longitudinal momentum fraction $z_h\approx 0.6$. At larger values of $z_h$, the displayed ratio falls off until $z_h\approx 0.8$ and then rises again.

To better understand the peak structure at intermediate values of $z_h$ as observed in Fig.~\ref{fig:NLORESTHR}, we consider the quark and gluon contributions separately for the same kinematics. In Fig.~\ref{fig:RESQG}, we show the enhancement due to the threshold resummation separately for quarks (blue) and gluons (red). For the quark, we show the result for $u$ quarks as an example. We define the quark and gluon contributions according to the resummed and matched jet functions ${\cal G}_c^h$ as given in Eqs.~(\ref{eq:NLOfact}) and~(\ref{eq:refactorize}). As expected, the numerical size of the enhancement is significantly larger for the gluon case due to larger color factors and the steeply falling gluon fragmentation function for $z_h\to 1$. Instead, for quarks the enhancement is moderate, and only close to the hadronic threshold do we observe a steep rise due to the resummation. However, when the two channels are combined, one finds that the gluon contribution is numerically significantly smaller than the quark contribution close to the hadronic threshold. This is mainly due to the fact that the gluon fragmentation function vanishes faster for $z_h\to 1$ than the quark fragmentation functions. Therefore, the large enhancement of the gluon channel seen in Fig.~\ref{fig:RESQG} does not directly translate to the entire cross section as shown in Fig.~\ref{fig:NLORESTHR}. Instead, at very large $z_h$, the cross section is dominated by the quark threshold enhancement. At intermediate $z_h$ the large enhancement of the gluon channel is still visible which leads to the peak around $z_h\approx 0.6$.

\begin{figure}[t]
\centering
\includegraphics[width=0.7\textwidth]{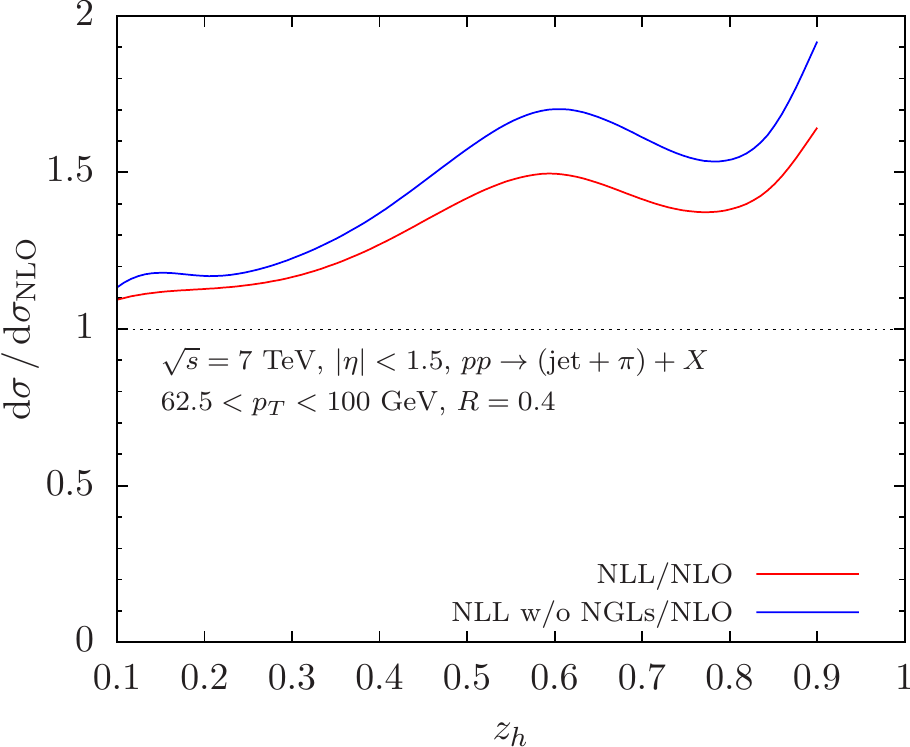}  
\caption{Ratio of the threshold resummed cross section and the NLO with (red) and without (blue) NGLs. The kinematics are the same as in Fig.~\ref{fig:NLORESTHR}.~\label{fig:RESNGL}} 
\end{figure}

Next, we consider the scale dependence of the NLO and the threshold resummed cross sections. In Fig.~\ref{fig:NLORESscale}, we show the QCD scale uncertainty bands for the NLO (hatched red band) and the threshold resummed calculation (green band). We vary both the hard scale $\mu_H\sim p_T$ and the jet scale $\mu_{\cal H}\sim\mu_{\cal G}\sim p_T R$ by factors of 2 around their central values and take the envelope. Here we show the cross section normalized to the inclusive jet cross section $\sigma_{\rm incl}$, and the kinematics are the same as in Fig.~\ref{fig:NLORESTHR} above. Note that the inclusive jet cross section differential in the jet's transverse momentum and rapidity can be obtained from Eq.~(\ref{eq:fact1}) by replacing the jet functions ${\cal G}_c^h$ with the semi-inclusive jet functions of~\cite{Kaufmann:2015hma,Kang:2016ehg,Dai:2016hzf}. We observe a significant reduction of the QCD scale dependence when threshold resummation is included.

Finally, we study the impact of the NGLs which are resummed in the large-$N_c$ approximation. In Fig.~\ref{fig:RESNGL}, we show the ratio of the threshold resummed cross section and the NLO one in blue when the NGLs are not included. For comparison, we also show the result of the resummation including NGLs (red) and we use the same kinematics as in Fig.~\ref{fig:NLORESTHR} above. We observe that the NGLs lead to a reduction of the cross section in the intermediate and large $z_h$ region as expected due to the additional soft emissions inside the jet. In the intermediate $z_h$ region the impact of the NGLs is somewhat larger since the gluon contribution dominates, as discussed above. Overall the impact of the NGLs is of the order of 10-15\% for the kinematics considered here.

\section{Conclusions}
\label{sec:conclusions}

In this work we considered the longitudinal momentum distribution of hadrons inside a reconstructed jet in proton-proton collisions. In the kinematic region where the momentum fraction of the identified hadron relative to the total jet momentum approaches unity, large logarithmic corrections arise which need to be resummed to all orders. These logarithmic corrections arise at the parton level and are further enhanced by the steeply falling fragmentation functions even away from the hadronic threshold. In this work, we developed a formalism within Soft Collinear Effective Theory that allows for the joint resummation of these threshold logarithms as well as of logarithms of the jet radius. The resummation was carried out in Mellin moment space and we identified the relation to resummed expressions in the traditional perturbative QCD literature. The resummation of logarithms of the jet radius was achieved through two DGLAP evolution equations of the jet function and the collinear fragmentation function. In order to achieve full next-to-leading logarithmic NLL$'$ accuracy, we included non-global logarithms in the leading-color approximation which arise due to configurations where a soft gluon is emitted into the jet from out-of-jet radiation. The Mellin inverse transformation was performed numerically using the so-called minimal prescription. Overall, we found that the threshold resummation leads to a sizable enhancement of the cross section compared to NLO and to a reduced QCD scale uncertainty. We also observed an intricate interplay between the resummations for the quark and gluon contributions to the cross section, leading to a pronounced peak structure of the resummed cross section at intermediate values of $z_h$. We found that the contribution due to non-global logarithms reduces the cross section in the threshold region. Our findings suggest that it is important to include threshold resummation for the extraction of fragmentation functions when hadron-in-jet data is included in global analyses. In addition, we expect that our results can shed new light on the non-trivial modification pattern of the jet fragmentation function in heavy-ion collisions.

\acknowledgments We would like to thank Wouter Waalewijn for helpful discussions. X.L. is supported by the National Natural Science Foundation of China under Grant No. 11775023 and the Fundamental Research Funds for the Central Universities. F.R. is supported by the Department of Energy under Contract No. DE-AC0205CH11231, the LDRD Program of LBNL and by the National Science Foundation under Grant No. ACI-1550228 within the JETSCAPE Collaboration. This work has been supported in part by the Bundesministerium f\"{u}r Bildung und Forschung (BMBF) under grant numbers 05P15VTCA1 and 05P18VTCA1. A.M. thanks the University of T\"ubingen for support where part of this work was done.

\bibliographystyle{JHEP}
\bibliography{bibliography}

\end{document}